\documentclass[journal]{IEEEtran}
   \usepackage[pdftex]{graphicx}
  \DeclareGraphicsExtensions{.pdf,.jpeg,.png}
\usepackage[]{units}
\usepackage[T1]{fontenc} 
\usepackage{amsmath}
\usepackage{amssymb}
\usepackage{bm} 
\usepackage{eurosym}
\usepackage[super]{nth} 
\usepackage{multirow}
\usepackage{color}
\usepackage[caption=false,font=footnotesize]{subfig}
\usepackage{enumitem}
\usepackage{cite}

\usepackage{longtable} 
\usepackage{eurosym}
\usepackage{comment}
\usepackage{float}

\usepackage[table]{xcolor}

\renewcommand{\baselinestretch}{0.96}

\begin{document}
%
\title{Efficient Database Generation for Data-driven Security Assessment of Power Systems}
%
%
%

\author{Florian~Thams,~\IEEEmembership{Student~Member,~IEEE,}
    Andreas~Venzke,~\IEEEmembership{Student~Member,~IEEE,}
        Robert~Eriksson,~\IEEEmembership{Senior~Member,~IEEE,}
        Spyros~Chatzivasileiadis,~\IEEEmembership{Senior~Member,~IEEE}
\thanks{F. Thams, A. Venzke, and S. Chatzivasileiadis are with the Center for Electric Power and Energy (CEE), Technical University of Denmark (DTU), Kgs. Lyngby, Denmark. E-mail: \{fltha, andven, spchatz\}  @elektro.dtu.dk.}
\thanks{R. Eriksson is with the Department of Market and System Development, Svenska kraftn\"at, Sundbyberg, Sweden. E-mail: robert.eriksson@svk.se}}

%
%

\markboth{Accepted at IEEE Transactions on Power Systems}%
{Thams \MakeLowercase{\textit{et al.}}: Efficient Database Generation for Data-driven Security Assessment of Power Systems}
%



\maketitle

\begin{abstract}
Power system security assessment methods require large datasets of operating points to train or test their performance. As historical data often contain limited number of abnormal situations, simulation data are necessary to accurately determine the security boundary. Generating such a database is an extremely demanding task, which becomes intractable even for small system sizes. This paper proposes a modular and highly scalable algorithm for computationally efficient database generation. Using convex relaxation techniques and complex network theory, we discard large infeasible regions and drastically reduce the search space. We explore the remaining space by a highly parallelizable algorithm and substantially decrease computation time. Our method accommodates numerous definitions of power system security. Here we focus on the combination of N-k security and small-signal stability. Demonstrating our algorithm on IEEE 14-bus and NESTA  162-bus systems, we show how it outperforms existing approaches requiring less than 10\% of the time other methods require.
\end{abstract}


%
\IEEEpeerreviewmaketitle

\section{Introduction}
%
%
%
%

\IEEEPARstart{S}{ecurity} assessment is a fundamental function for both short-term and long-term power system operation. Operators need to eliminate any probability of system failure on a sub-hourly basis, and need to guarantee the security of supply in the long-term, having the required infrastructure and operating practices in place. All these functions require the assessment of thousands of possibilities with respect to load patterns, system topology, power generation, and the associated uncertainty which is taking up a more profound role with the increased integration of renewable energy sources (RES). Millions of possible operating points violate operating constraints and lead to an insecure system, while millions satisfy all limitations and ensure safe operation. For systems exceeding the size of a few buses it is impossible to assess the total number of operating points, as the problem complexity explodes. Therefore, computationally efficient methods are necessary to perform a fast and accurate dynamic security assessment.
%

Numerous approaches exist in the literature proposing methods to assess or predict different types of instability, e.g. transient, small-signal, or voltage instability. Recently, with the abundance of data from sensors, such as smart meters and phasor measurement units, machine learning approaches have emerged showing promising results in tackling this problem \cite{Konstantelos2016, Preece2016, IREP2017, Lejla_PSCC}. Due to the high reliability of the power system operation, however, historical data are not sufficient to train such techniques, as information related to the security boundary or insecure regions is often missing. For that, simulation data are necessary. 

This paper deals with the fundamental problem that most of the dynamic security assessment (DSA) methods are confronted with before the implementation of any algorithm: the generation of the necessary dataset which is required for the development of dynamic security classification approaches. With this work we aim to propose a modular and scalable algorithm that can map the secure and insecure regions, and identify the security boundaries of large systems in a computationally efficient manner. 

There are two main challenges with the generation of such a database. First, the problem size. It is computationally impossible to assess all possible operating points for systems exceeding a few tens of buses. Second, the information quality. Dynamic security assessment is a non-convex and highly nonlinear problem. Generating an information-rich and not too large dataset can lead to algorithms that can be trained faster and achieve higher prediction accuracy.

The efforts to develop a systematic and computationally efficient methodology to generate the required database have been limited up to date. In \cite{Wehenkel1994, Hatziargyriou1994} re-sampling techniques based on post-rule validation were used to enrich the database with samples close to the boundary. Genc et al. \cite{Genc2010} propose to enrich the database iteratively with additional points close to the security boundary by adding operating points at half the distance of the already existing operating points at the stability boundary. In \cite{Liu2013b,Liu2014,Krishnan2011,Preece2016,Hamon2016}, the authors propose to use importance sampling methods based on Monte-Carlo variance reduction (MCVR) technique, introducing a bias in the sampling process such that the representation of rare events increases in the assessment phase. In \cite{Sun2016}, the authors propose a composite modelling approach using high dimensional historical data.

This work leverages advancements in several different fields to propose a highly scalable, modular, and computationally efficient method. Using properties derived from convex relaxation techniques applied on power systems, we drastically reduce the search space. Applying complex network theory approaches, we identify the most critical contingencies boosting the efficiency of our search algorithms. Based on steepest descent methods, we design the exploration algorithm in a highly parallelizable fashion, and exploit parallel computing to reduce computation time. Compared with existing approaches, our method achieves a speed-up of 10 to 20 times, requiring less than 10\% of the time other approaches need to achieve the same results.

The contributions of this work are the following:
\begin{itemize}
    \item We propose a computationally efficient and highly scalable method to generate the required datasets for the training or testing of dynamic security assessment methods. Our approach requires less than 10\% of the time existing methods need for results of similar quality.
    \item Our method is modular and can accommodate several types of security boundaries, including transient stability and voltage stability. In this paper, we demonstrate our approach considering the combination of N-k security and small signal stability.
    \item Besides the database generation, the methodology we propose can be easily employed in real-time operation, where computationally efficient techniques are sought to explore the security region in case of contingencies around the current operating point.
    \item In case studies we demonstrate the importance of a high quality database to achieve the best possible results in a data-driven security assessment. Given equal computation time, training machine learning algorithms with the database generated by our method clearly outperforms other approaches.
\end{itemize}
The remainder of this paper is organized as follows: First, a set of terms are defined in Section~\ref{sec:definitions}. In Section \ref{sec:Challenges}, we describe the challenges of the database generation for data-driven security analysis. Section \ref{sec:Methodology} provides an overview of the methodology, which we detail in the two subsequent sections. Section~\ref{sec:reducingsearchspace} describes how we reduce the search space, while Section~\ref{sec:directedwalks} describes the highly parallelizable exploration of the remaining space. We demonstrate our methods in Section~\ref{sec:Case2}. Section~\ref{sec:conclusion} concludes the paper.
\section{Definitions}
\label{sec:definitions}
\subsubsection{Security boundary}
the boundary $\gamma$ dividing the secure from the insecure region; (a) can correspond to a specific stability boundary, e.g. small-signal stability or voltage stability, (b) can represent a specific stability margin, i.e. all operating points not satisfying the stability margin belong to the insecure region, (c) can be a combination of security indices, e.g. the intersection of operating points that are both N-1 secure and small-signal stable. Note that our proposed method can apply to any security boundary the user needs to consider.
\subsubsection{HIC -- High Information Content}
the set $\Omega$ of operating points in the vicinity of the security boundary $\gamma$, see \eqref{eq:high_inf} \cite{Krishnan2011}. This is the search space of high interest for our methods as it separates the secure from insecure regions.
\subsubsection{DW -- Directed Walk} we use this term to denote the steepest descent path our algorithm follows, starting from a given initialization point, in order to arrive close to the security boundary.
\section{Challenges of the Database Generation} \label{sec:Challenges}
Determining the secure region of a power system is an NP-hard problem. In an ideal situation, in order to accurately determine the non-convex secure region we need to discretize the whole space of operating points with an as small interval as possible, and perform a security assessment for each of those points.
For a given system topology, this set consists primarily of all possible combinations of generator and load setpoints (Note that if the system includes tap-changing transformers, and other controllable devices, the number of credible points increases geometrically).  Thus, in a classical brute force approach, the number of points to be considered is given by:
 \begin{align}
     \left|\Psi \right| = \Lambda \cdot \prod_{i=1}^{N_G-1} \Big( \frac{P_{i}^{max}-P_{i}^{min}}{\alpha}+1 \Big), \label{eq:numberOP}
 \end{align}
 where $N_G$ is the number of generators $i$, $P_{i}^{max}$ and $P_{i}^{min}$ is their maximum and minimum capacity, $\alpha$ is the chosen discretization interval between the generation setpoints, and $\Lambda$ represents the number of different load profiles.
 For example, for the IEEE 14 bus system \cite{Zimmerman2011}
 with 5 generators and a discretization interval of $\alpha = \unit[1]{MW}$, a classical brute force approach requires
 $\left|\Psi \right| \approx 2.5 \cdot 10^6$
 operating points to be assessed for a \emph{single load profile}. It can be easily seen that security assessment of large systems can very fast result in an intractable problem. For example, in the NESTA 162 bus system \cite{Coffrin2014}, a brute force approach would require the analysis of $7\cdot 10^{29}$ points. It becomes clear that the efficient database generation is one of the major challenges for the implementation of data-driven tools in power system security analysis. In this effort, we need to balance the trade-off between two partially contradicting goals: keep the database as small as possible to minimize computational effort, but contain enough data to determine the security boundary as accurately as possible.
\begin{figure}[!b]
\vspace{-3ex}
\centering
\includegraphics[width=2.3in]{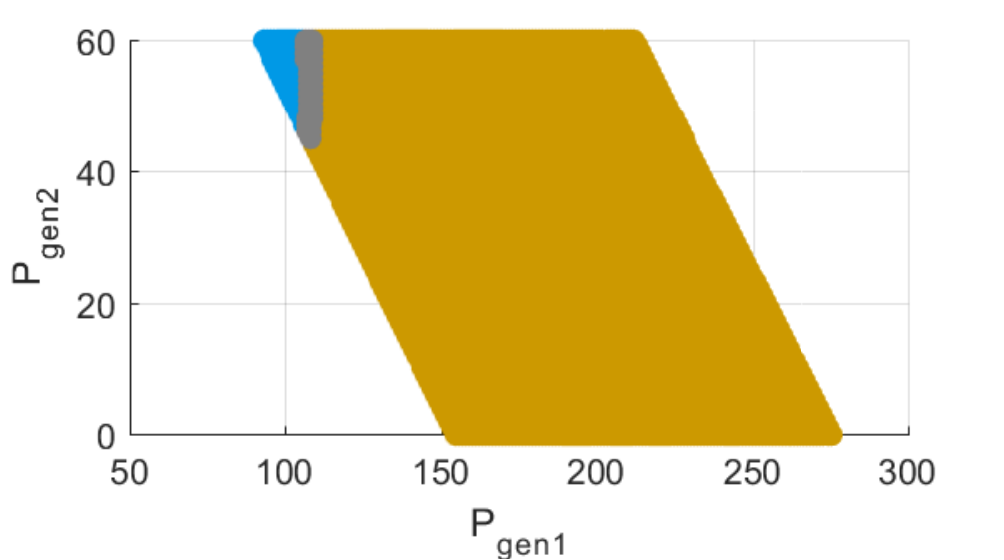}
\vspace{-2ex}
\caption{Scatter plot of all possible operating points of two generators for a certain load profile. Operating points fulfilling the stability margin and outside the high information content (HIC) region ($\gamma_k > 3.25\%$) are marked in blue, those not fulfilling the stability margin and outside HIC ($\gamma_k < 2.75\%$) are marked in yellow. Operating points located in the HIC region ($2.75\% < \gamma_x < 3.25\%$) are marked in grey.}
\label{fig_lineflow}
\vspace{-3ex}
\end{figure}
%
To better illustrate our approach, in Fig.~\ref{fig_lineflow} we show all possible operating points of two generators for a certain load profile in a system. 
Focusing on small-signal stability here, we define the security boundary $\gamma$ as a certain level of minimum damping ratio, which corresponds to our stability margin. All safe operating points, with a damping ratio below $\gamma$, are plotted in blue, while operating points that do not fulfill the stability margin are plotted in yellow. From Fig.\ref{fig_lineflow}, it is obvious that if we are able to assess all points close to $\gamma$, it is easy to classify the rest of the points. By that, the size of the required database can be significantly reduced.
In the remainder of this paper, we will call the set of operating points in the vicinity of $\gamma$ as the set of \emph{high information content (HIC)}, defined as follows:
 \begin{align}
   \label{eq:high_inf}
    \Omega =\{OP_k\in \Psi\mid  \gamma-\mu < \gamma_k < \gamma +\mu\},
 \end{align}
with $\gamma_k$ denoting the value of the chosen stability margin for operating point $OP_k$ and $\mu$ representing an appropriate small value to let $\left|\Omega\right|$ be large enough to describe the desired security boundary with sufficient accuracy. The value of $\mu$ depends on the chosen discretization interval in the vicinity of the boundary. In Fig. \ref{fig_lineflow}, the HIC set, i.e. all points $OP_k\in \Omega$, is visualized as the grey area surrounding $\gamma$. In this small example, we were able to assess all possible operating points and accurately determine the HIC area. For large systems this is obviously not possible. As a result, in the general case, the main challenge is to find the points $OP_k$ which belong to the HIC area $\left|\Omega \right|$.

To put the difference between $\left|\Psi \right|$ and $\left|\Omega \right|$ in perspective: for the small signal stability analysis of the IEEE 14 bus system, the classical brute force approach requires the analysis of $\left|\Psi \right| \approx \unit[2.5 \cdot 10^6]{}$ operating points (OPs) for a single load profile. By assuming a required damping ratio of $\gamma = \unit[3]{\%}$,
and $\mu = \unit[0.25]{\%}$, the HIC set, defined as $\Omega =\{OP_k\in \Psi\mid  \unit[2.75]{\%} < \gamma_k< \unit[3.25]{\%}\}$, reduces the analysis to only $1457$ points (here, $\gamma$ refers to the damping ratio of the lowest damped eigenvalue). In other words, by assessing only $\unit[0.06]{\%}$ of all data points, we can accurately determine the whole secure region of this example.

%
This small amount of required operating points to assess has actually worked as an obstacle for one of the most popular approaches in previous works: importance sampling. Importance sampling re-orients the sampling process towards a desired region of interest in the sampling space, while also preserving the probability distribution. Thus, it requires that the initial sampling points include sufficient knowledge about the desired region of interest. However, the smaller the proportion of the region of interest is in respect to the entire multi-dimensional space, the larger the initial sample size needs to be to include a sufficient number of points within the desired region of interest. Therefore, the use of expert knowledge \cite{Wehenkel1994, Hatziargyriou1994, Krishnan2011}, regression models \cite{Preece2016} or linear sensitivities \cite{Krishnan2011} are proposed to determine the desired region and reduce the search space. However, since this search space reduction is based on a limited initial sample size, it entails the risk of either missing regions of interest not represented in the initial sample or requires a large initial sample which increases computational burden. Furthermore, previous works \cite{Krishnan2011,Genc2010} often use expert knowledge to reduce the burden of the N-1 security assessment to a few critical contingencies.
Our proposed method does not require expert knowledge and avoids potential biases by not separating the knowledge extraction from the sampling procedure. Still, if expert knowledge of e.g. a preferred search region or the most critical contingencies is available, our method can easily integrate it and benefit from it.
\section{Methodology} \label{sec:Methodology}
We divide the proposed methodology in two main parts. First, the search space reduction by the elimination of a large number of infeasible (and insecure) operating points. Second, the directed walks: a steepest-descent based algorithm to explore the search space and accurately determine the security boundary. During the search space reduction, we exploit properties of convex relaxation techniques to discard large infeasible regions. In order to reduce the problem complexity, we employ complex network theory approaches which allow us to identify the most critical contingencies. Finally, designing the directed walks as a highly parallelizable algorithm, we use parallel computing capabilities to drastically reduce the computation time. 

The different parts are described in detail in the following sections. Our algorithm starts by uniformly sampling the search space using the Latin Hypercube Sampling (LHS) method to generate initialization points for the subsequent steps (Section~\ref{seq:initPoints}). Following that, we propose a convex grid pruning algorithm, which also considers contingency constraints, to discard infeasible regions and reduce the search space (Section~\ref{sec:GridPruning}). In Section~\ref{seq:VulnerableCont.}, we leverage complex network theory approaches to identify the most critical contingencies. The identified contingency set is crucial both for the grid pruning algorithm, and for subsequent steps within the Directed Walks. After resampling the now significantly reduced search space, we use these samples as initialization points for the Directed Walk (DW) algorithm, described in Section~\ref{sec:directedwalks}. In order to achieve an efficient database generation, the goal of the algorithm is to traverse as fast as possible large parts of the feasible (or infeasible) region, while carrying out a high number of evaluations inside the HIC region. This allows the algorithm to focus on the most relevant areas of the search space in order to accurately determine the security boundary. The DWs are highly parallelizable, use a variable step size depending on their distance from the security boundary, and follow the direction of the steepest descent. Defining the secure region as the N-2 secure \emph{and} small signal stable region in our case studies, we demonstrate how our method outperforms existing importance sampling approaches, achieving a 10 to 20 times speed-up.
\section{Reducing the Search Space}
\label{sec:reducingsearchspace}
\subsection{Choice of Initialization Points}\label{seq:initPoints}
An initial set of operating points is necessary to start our computation procedure. Using the Latin Hypercube Sampling (LHS), we sample the space of operating points to select the initialization points $\eta$. Besides the initialization points at the first stage, $\eta_1$, our method requires the finer selection of initialization samples, $\eta_2$ and $\eta_3$ during the reduction of the search space in two subsequent stages, as will be explained later. We use the same sampling procedure at all stages.
The Latin hypercube sampling (LHS) aims to achieve a uniform distribution of samples across the whole space. Dividing each dimension in subsections, LHS selects only one sample from each subsection of each parameter, while, at the same time, it maximizes the minimum distance between the samples \cite{Preece2016}. An even distribution of the initialization points over the multi-dimensional space is of high importance in order to increase the probability that our method does not miss any infeasible region or any HIC region. The number of initialization points $|\eta|$ is a tuning factor which depends on the specific system under investigation. In general, quite sparse discretization intervals are used for the search space reduction procedures in the first two stages of our approach, $\eta_{1-2}$, while a more dense discretization interval is used for the directed walks at a later stage, $\eta_{3}$. Suitable values are discussed in the case study. While LHS allows an even sampling, it is very computationally expensive for high-dimensional spaces and large numbers of initialization points. Thus, for larger systems there is a trade-off between initial sampling and computation time that needs to be considered.
%
\subsection{Grid Pruning Algorithm For Search Space Reduction}\label{sec:GridPruning}
Given the $\eta_{1}$ initialization points from the first stage, the aim of this stage is to reduce the search space by eliminating infeasible operating regions. For that, we use a grid pruning algorithm which relies on the concept of convex relaxations. The algorithm is inspired by \cite{Molzahn2016}, where it was developed to compute the feasible space of small AC optimal power flow (OPF) problems. In this work, we introduce a grid pruning algorithm which determines infeasible operating regions considering not only the intact system but also all N-1 contingencies.

Convex relaxations have been recently proposed to relax the non-convex AC-OPF to a semidefinite program \cite{Lavaei2012}. A corollary of that method is that the resulting semidefinite relaxation provides an infeasibility certificate: If an initialization point is infeasible for the semidefinite relaxation, it is guaranteed to be infeasible for the non-convex AC-OPF problem. This means that for that initialization point there does not exist a power flow solution which complies with all operational constraints (i.e. voltage limits, active / reactive power limits). A feasible power flow solution is a basic requirement for a security assessment using any stability metric. This property of the semidefinite relaxation is used in our grid pruning algorithm.

The semidefinite relaxation introduces the matrix variable $W$ to represent the product of real and imaginary parts of the complex bus voltages (for more details the interested reader is referred to \cite{Lavaei2012,Molzahn2013}). Defining our notation, the investigated power grid consists of $\mathcal{N}$ buses, where $\mathcal{G}$ is the set of generator buses. We consider a set of line outages $\mathcal{C}$, where the first entry $\{0\}$ of set $\mathcal{C}$ corresponds to the intact system state. The following auxiliary variables are introduced for each bus $i \in \mathcal{N}$ and outage $c \in \mathcal{C}$:
\begin{align}
	Y^c_i &:= e_i e_i^T Y^c & \label{aux1} \\
	\textbf{Y}^c_i & := \dfrac{1}{2} \begin{bmatrix} \Re \{Y^c_i + (Y^c_i)^T\} & \Im \{  (Y^c_i)^T - Y^c_i \} \\ \Im \{ Y^c_i -  (Y^c_i)^T\} & \Re \{Y^c_i +  (Y^c_i)^T\} \end{bmatrix} & \\
	\bar{\textbf{Y}}^c_i &:= \dfrac{-1}{2} \begin{bmatrix} \Im \{Y^c_i + (Y^c_i)^T\} & \Re \{ Y^c_i - (Y^c_i)^T \} \\ \Re \{ (Y^c_i)^T - Y^c_i\} & \Im \{Y^c_i + (Y^c_i)^T\} \end{bmatrix} \\
	M_i &:= \begin{bmatrix} e_i e_i^T & 0 \\ 0 & e_i e_i^T \end{bmatrix}
\end{align}
Matrix $Y^c$ denotes the bus admittance matrix of the power grid for outage $c$, and $e_i$ is the i-th basis vector. The operators $\Re$ and $\Im$ denote the real and imaginary parts of the matrix. The initialization points $\eta_1$ from stage A (see Section~\ref{seq:initPoints}) correspond to both feasible and infeasible operating points for the AC optimal power flow problem. Given a set-point $P^*$ for the generation dispatch (corresponding to initialization point $\eta_{1}^{*}$), \eqref{CR_obj} -- \eqref{LINK_SC2} compute the minimum distance from $P^*$ to the closest feasible generation dispatch. Obviously, if $P^*$ is a feasible generation dispatch, the minimum distance is zero.
\begin{align}
    \min_{W^c} \, & \sqrt{ \sum_{i \in \mathcal{G}} ( \text{Tr} \{ \textbf{Y}^0_i W^0\} + P_{D_i} - P^*_i )^2 } \label{CR_obj} \\
    \text{s.t.} \, &	\underline{P}_{G_i}  \leq \text{Tr} \{ \textbf{Y}^c_i W^c\} + P_{D_i} \leq \overline{P}_{G_i}  \quad \forall i \in \mathcal{N} \, \forall c \in \mathcal{C}  \label{PBal}                \\
&	\underline{Q}_{G_i}   \leq \text{Tr} \{ \bar{\textbf{Y}}^c_i W^c\} + Q_{D_i} \leq \overline{Q}_{G_i} \quad \forall i \in \mathcal{N} \, \forall c \in \mathcal{C}  \label{QBal}          \\
&	\underline{V}_i^2 \leq \text{Tr} \{ M_i W^c\} \leq \overline{V}_i^2 \quad \forall i \in \mathcal{N} \, \forall c \in \mathcal{C}   \label{VCon} \\
&  W^c \succeq 0 \quad \forall c \in \mathcal{C} \label{SDP} \\
& \text{Tr} \{ \textbf{Y}^c_i W^c\} = \text{Tr} \{ \textbf{Y}^0_i W^0\} \quad \forall i \in \mathcal{G} \backslash \{\text{slack}\} \, \forall c \in \mathcal{C} \label{LINK_SC1} \\
& \text{Tr} \{ M_i W^c\} = \text{Tr} \{ M_i W^0\} \quad \forall i \in \mathcal{G} \, \forall c \in \mathcal{C} \label{LINK_SC2}
\end{align}
The matrix variable $W^0$ refers to the intact system state with admittance matrix $\textbf{Y}^0$. The objective function \eqref{CR_obj} minimizes the distance of the active generation dispatch from the set-point $P^*$. The operator $Tr\{\}$ denotes the trace of a matrix. For each outage $c \in \mathcal{C}$, one matrix variable $W^c$ is introduced which is constrained to be positive semidefinite in \eqref{SDP}. The terms $\underline{P}_{G_i}$, $\overline{P}_{G_i}$, $\underline{Q}_{G_i}$, $\overline{Q}_{G_i}$ in the nodal active and reactive power balance \eqref{PBal} and \eqref{QBal} are the maximum and minimum active and reactive power limits of the generator at bus $i$, respectively. The active and reactive power demand at bus $i$ is denoted with the terms $P_{D_i}$ and $Q_{D_i}$. The bus voltages at each bus $i$ are constrained by upper and lower bounds  $\overline{V}_i$ and $\underline{V}_i$ in \eqref{VCon}. In case of an outage, the generator active power and voltage set-points remain fixed \eqref{LINK_SC1} -- \eqref{LINK_SC2}, as traditional N-1 (and \mbox{N-k}) calculations do not consider corrective control. To reduce the computational complexity of the semidefinite constraint \eqref{SDP}, we apply a chordal decomposition according to \cite{Molzahn2013} and enforce positive semidefiniteness only for the maximum cliques of matrix $W^c$. To obtain an objective function linear in $W^0$, we introduce the auxiliary variable $R$ and replace \eqref{CR_obj} with:
 \begin{align}
    \min_{W^c,R} \quad & R \label{Cr_obj}\\
     \text{s.t.} & \sqrt{ \sum_{i \in \mathcal{G}} ( \text{Tr} \{ \textbf{Y}_i^0 W^0\} + P_{D_i} - P^*_i )^2 } \leq R \label{SOC_D}
 \end{align}
 The convex optimization problem \eqref{PBal} -- \eqref{SOC_D} guarantees that the hypersphere with radius $R$ around the operating point $P^*$ does not contain any points belonging to the non-convex feasible region, considering both the intact system state and the contingencies in set $\mathcal{C}$. Note that the obtained closest generation dispatch $P_i^0 = \text{Tr} \{ \textbf{Y}_i^0 W^0\} + P_{D_i}$ is feasible in the relaxation but not necessarily in the non-convex problem. Hence, with $R$ we obtain a lower bound of the distance to the closest feasible generation dispatch in the non-convex problem. 

\begin{figure}[!t]
\centering
\includegraphics[width=2.7in]{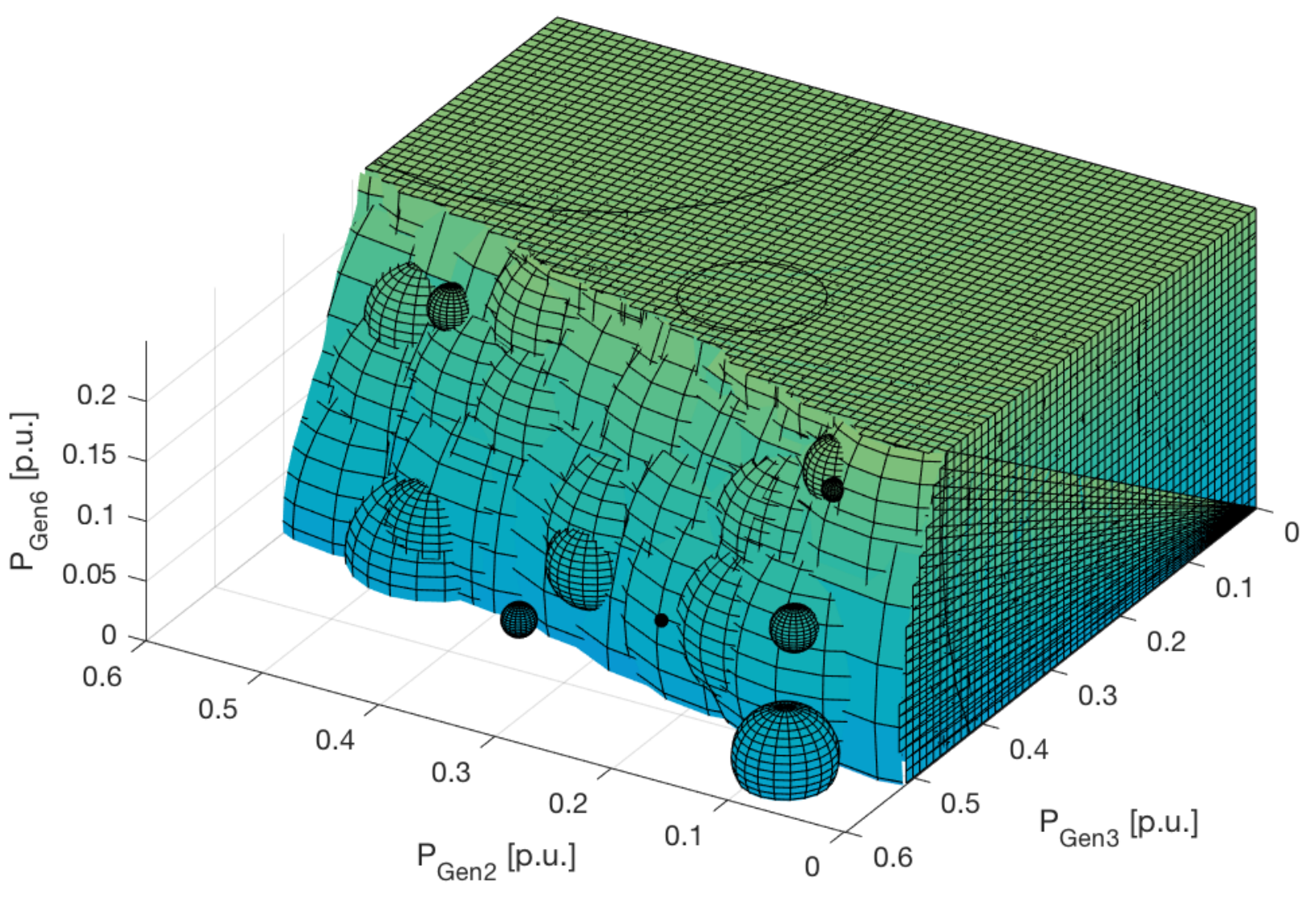}
\caption{Search space reduction obtained by the proposed grid pruning algorithm for the IEEE 14 bus system. Operating points within the structure formed by superimposed spheres are infeasible considering N-1 security.}
\label{fig_LHS}
\end{figure}

In a procedure similar to \cite{Venzke_SDP_SCOPF_PSCC2018}, we apply an iterative algorithm for the grid pruning: First, given $\eta_{1}$ initialization points, we solve \eqref{PBal} -- \eqref{SOC_D} without considering contingencies, i.e. $\mathcal{C} = \{ 0 \}$. Using the determined hyperspheres, we eliminate the infeasible operating regions and, using LHS, we resample the reduced search space to select a set of initialization points $\eta_2$.  In the next stage, given $\eta_{2}$, we determine the five most critical contingencies (see section \ref{seq:VulnerableCont.} for more details) and resolve \eqref{PBal} -- \eqref{SOC_D}. We remove all resulting infeasible regions from the set $\eta_2$, and using LHS we resample the remaining feasible region to determine the initialization set $\eta_{3}$. The number of considered contingencies is a trade-off between the amount of obtained infeasible points and the required computational time to solve the semidefinite relaxation. The number of initialization points $\eta_{1-2}$ should be chosen to minimize the overlapping of the hyperspheres while maximizing the search space reduction. As all $\eta_{2}$ points within the infeasible region are immediately discarded, a value $\eta_{2}>\eta_{1}$ is required in order to obtain a smaller distance between the initialization points and, thus more points within the feasible region than before. However, the more the resulting hyperspheres are overlapping (as. e.g. visualized in Fig.~\ref{fig_LHS}), the less information every point is providing and the less computationally efficient the grid pruning is. Thus, the choice depends also on the system size, as the same number of initialization points will lead to different distances between the points depending on the system size. Finally, $\eta_{3}$ needs to be chosen large enough to obtain sufficient initialization points for the directed walks but not too large in order to avoid too many duplicates being created during the walks.

As an example of the search space reduction, we consider the IEEE 14 bus system in a scenario where all but three generators ($P_{gen 2-4}$) are fixed to specific values. Considering the five most critical contingencies, and using our proposed convex grid pruning algorithm, the search space is reduced by $\unit[65.34]{\%}$. This is visualized in Fig.~\ref{fig_LHS}; the colored area shows the discarded regions of infeasible points as determined by the superposition of the spheres.  
\subsection{Determining the Most Critical Contingencies} \label{seq:VulnerableCont.}
From the definition of N-1 security criterion it follows that a single contingency suffices to classify an operating point as infeasible.
Most of the unsafe operating points, however, belong to the infeasible regions of several contingencies. As a result, focusing only on a limited number of \emph{critical} contingencies, we can accurately determine a large part of the N-k insecure region, thus reducing the search space without the need to carry out a redundant number of computations for the whole contingency set. This drastically decreases the computation time.

The goal of this section is to propose a methodology that determines the most critical contingencies, which can then be used both in the convex grid pruning algorithm \eqref{PBal} -- \eqref{SOC_D}, and in the step direction of the DWs in Section~\ref{seq:DirectionDet.}.
While classical N-1 (and N-k) analyses are computationally demanding, recent approaches based on complex network theory showed promising results while requiring a fraction of that time. Refs. \cite{Albert2004}, \cite{Fang2016} propose fast identification of vulnerable lines and nodes, using concepts such as the (extended) betweenness or the centrality index.

%
The centrality index used in \cite{Fang2016}, and first proposed for power systems in \cite{Dwivedi2010, Dwivedi2013}, is based on a classical optimization problem in complex network theory, known as maximum flow problem. The index refers to the portion of the flow passing through a specific edge in the network. Components with higher centrality have a higher impact on the vulnerability of the system, and thus have higher probability to be critical contingencies.
%

Similar to \cite{Fang2016}, we adopt an improved max-flow formulation for the power system problem which includes vertex weights, and extends the graph with a single global source and a single global sink node. The improved formulation accounts for the net load and generation injections at every vertex, avoids line capacity violations resulting from the superposition of different source-sink combinations, and decreases computation time. Contrary to \cite{Fang2016}, however, we use a modified definition of the centrality index.
While Fang et al. \cite{Fang2016} analyze the most critical contingencies for all generation and demand patterns, we are interested in the most critical contingency for every \emph{specific} load and generation profile, i.e. for every operating point $OP_k$. Thus, for each operating point $OP_k$ we define the centrality index as:
 \begin{align}\label{eq:centrality}
     C_{ij}^{(k)}= f_{ij,actual}^{(k)} / f_{max}^{(k)} \quad \forall i,j \in \mathcal{N},
 \end{align}
where $f_{ij,actual}^{(k)}$, are the actual flows for that operating point $OP_k$, and $f_{max}^{(k)}$ represents the maximum possible flow between global source and global sink node for the same case. Thus, at every operating point $OP_k$, the lines are ranked according to their contribution to the maximum flow in the system. The higher their centrality index is, the more vulnerable becomes the system in case they fail, and as a result they are placed higher in the list of most critical contingencies.

The case study includes a brief discussion about the performance of this vulnerability assessment for the investigated systems. Despite the drastic decrease in computation time and its general good performance, the proposed approach still includes approximations. As we will see in Sections~\ref{seq:FinalEval}--\ref{seq:Round2}, we take all necessary steps to ensure that we have avoided any possible misclassification.
\begin{figure*}[!t]
\vspace{-2ex}
\centering
\includegraphics[width=6in]{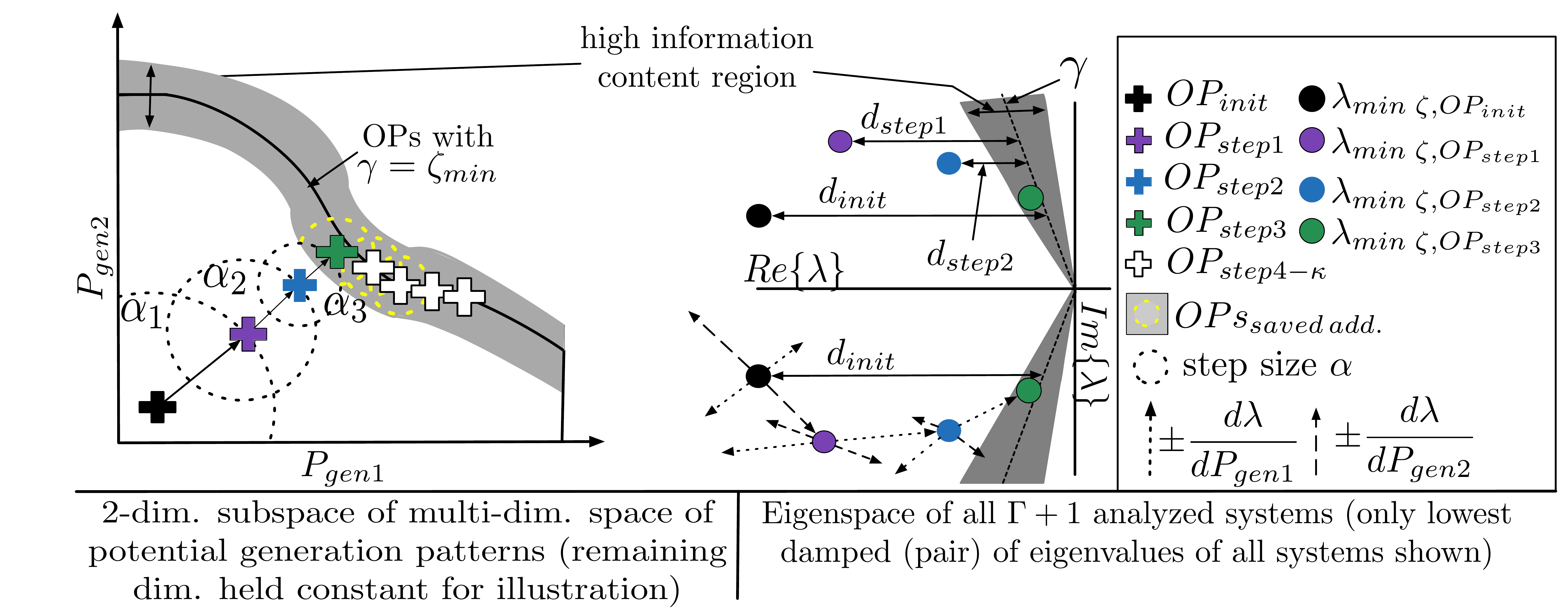}
\vspace{-2ex}
\caption{Illustration of the Directed Walk (DW) through a two dimensional space using varying step sizes, $\alpha_i$, following the steepest descent of distance, $d$.
}
\label{fig_direction}
\vspace{-3ex}
\end{figure*}
\section{Directed Walks}
\label{sec:directedwalks}
\subsubsection{Variable Step Size}\label{seq:StepSize.}
As mentioned in Section~\ref{sec:Challenges}, to achieve an efficient database generation our focus is to assess a sufficiently high number of points inside the HIC area, while traverse the rest of the space faster and with fewer evaluations. To do that, we propose to use a variable step size $\alpha$ depending on the distance $d$ of the operating point from the security boundary $\gamma$. The distance $d(OP_k)$ of the operating point under investigation is defined as:
\begin{align}\label{eq:distanceOP}
     d(OP_k) = \left|\gamma_k - \gamma \right|,
\end{align}
with $\gamma_k$ being the stability index value for operating point $OP_k$.
Then, for $OP_k$ we define the variable step size $\alpha_k$ as follows:
\begin{equation}
  \alpha_k=\left\{
  \begin{array}{@{}ll@{}}
    \epsilon_{1}\cdot P^{max}, & \text{if}\ d(OP_k) >d_1 \\
    \epsilon_{2}\cdot P^{max}, & \text{if}\ d_1 \geq d(OP_k) >d_2 \\
   \epsilon_{3}\cdot P^{max}, & \text{if}\ d_2 \geq d(OP_k) >d_3 \\
   \epsilon_{4}\cdot P^{max}, & \text{otherwise}
  \end{array} \quad , \right. 
  \label{eq:stepsize}
\end{equation}
where $P^{max}$ is the vector of generator maximum capacities, $\epsilon_{1-4}$ are scalars, and for distances $d_{1-3}$ holds $d_1>d_2>d_3$.
Since the system is highly nonlinear (in our case study for example we are searching for the minimum damping ratio considering N-1 security, i.e. $\left|\mathcal{C} \right|$ different nonlinear systems superimposed), the exact step size required to reach the HIC region cannot be constant or determined a-priori. Thus, the step size is gradually reduced as we approach the security boundary in order not to miss any points within the HIC region. This is illustrated in Fig. \ref{fig_direction}.

It follows that distances $d_{1-3}$ and the corresponding $\epsilon_{1-4}$ are tuning factors, to be chosen depending on the desired speed, granularity, precision and given system size. Factors found useful for given systems are discussed in the case study.
%
\subsubsection{Determining the Step Direction}\label{seq:DirectionDet.}
After identifying the step size, we need to determine the direction of the next step. Our goal is to traverse the feasible (or infeasible) region as fast as possible, and enter the HIC region. To do that, at every step we follow the steepest descent of the distance metric $d(OP_k)$, as shown in \eqref{eq:steepestdescent}.
\begin{align}\label{eq:steepestdescent}
    OP_{k+1} &= OP_k- \alpha_k \cdot \nabla d(OP_k)
\end{align}
where $\alpha_k$ is the step size for $OP_k$, defined in \eqref{eq:stepsize}, and $\nabla d(OP_k)$ is the gradient of $d(OP_k)$.
As the distance is a function of the chosen stability index, it is user specific and $\nabla d(OP_k)$ in the discrete space shall be determined by a suitable sensitivity measure, which differs for different stability indices. If the focus is on voltage stability for example, the associated margin sensitivities could be used \cite{Greene1997}. It is stressed that our method is suitable for any sensitivity capable of measuring the distance to the chosen stability index. In the case studies of this paper, we focus on small-signal stability and, as described in the next paragraph, we pick the damping ratio sensitivity as a suitable measure.

Normally, at every step $k$ we should measure distance $d$ for all N-1 (or N-k) contingencies, select the minimum of those distances and based on that, determine the next step size and direction. Having thousands of initialization points $\eta_3$ implies checking along all possible dimensions and N-1 contingencies at every step of thousands of directed walks. Beyond a certain system size, this becomes computationally intractable. Instead, we take advantage of the critical contingency identification procedure described in Section \ref{seq:VulnerableCont.}, and at every step we measure distance $d$ assuming the most critical contingency for $OP_k$. This reduces the required analysis from $\left|\mathcal{C} \right|$ systems to one system, which drastically decreases the computation time. Following steps in this procedure, as described later, ensure that this approximation is sufficient and there is an accurate detection of the security boundary as soon as we enter the HIC region. 
%
\subsubsection{Sensitivity Measure for Small-Signal Stability}
For small-signal stability, we determine the step direction by the sensitivity of the damping ratio, $\zeta$, of the system representing the most critical contingency, $c_c \in \mathcal{C}$. This requires to compute the eigenvalue sensitivity which, in turn, depends on the state matrix $\mathbf{A_{c_c}}$ (for more details about forming state matrix $\mathbf{A}$ the reader is referred to \cite{Kundur:1994tx}). Thus, the sensitivity of eigenvalue $\lambda_n$ to a system parameter $\rho_i$ is defined as
 \begin{align}\label{eq:eigenvalue_sens}
     \frac{\partial \lambda_n}{\partial \rho_i} = \frac{\mathbf{\psi}_n^T \frac{\partial \mathbf{A_{c_c}}}{\partial \rho_i} \mathbf{\phi_n}}{\mathbf{\psi}_n^T \mathbf{\phi}_n}.
 \end{align}
 $\mathbf{\psi}_n^T$ and $\mathbf{\phi}_n$ are the left and right eigenvectors, respectively, associated with eigenvalue $\lambda_n$ \cite{Kundur:1994tx}.
Defining $\lambda_n = \sigma_n +j \omega_n$, and with $\zeta=\frac{-\sigma_n}{\sqrt{\sigma_n^2 + \omega_n^2}}$, we can determine the damping ratio sensitivity, $\frac{\partial \zeta_n}{\partial \rho_i}$ as
  \begin{align}\label{eq:damp_sens}
     \frac{\partial \zeta_n}{\partial \rho_i} = \frac{\partial}{\partial \rho_i}  \left( \frac{-\sigma_n}{\sqrt{\sigma_n^2 + \omega_n^2}}\right) = \omega_n \frac{(\sigma_n\frac{\partial \omega_n}{\partial \rho_i}-\omega_n \frac{\partial \sigma_n}{\partial \rho_i})}{(\sigma_n^2 + \omega_n^2)^{\frac{3}{2}}}.
   \end{align}
Due to the fact that the computation of $\frac{\partial \mathbf{A_{c_c}}}{\partial \rho_i}$ is extremely demanding, it is usually more efficient to determine the damping ratio sensitivity of $\zeta$ to $\rho_i$ by a small perturbation of $\rho_i$. The whole process is illustrated in Fig. \ref{fig_direction}. The parameters $\rho_i$ correspond to the power dispatch of two generators. The DW is illustrated following the steepest descent of damping ratio considering the lowest damped eigenvalue of the system representing the most critical contingency $c_c \in \mathcal{C}$.
%
%
%
%
\subsubsection{Parallelization of the Directed Walks}
Directed Walks are easily parallelizable. In our case studies, we have used 80 cores of the DTU HPC cluster for this part of our simulations. To ensure an efficient parallelization and not allow individual processes take up unlimited time, we set a maximum number of steps of DWs, $\kappa_{max}$. The tuning of $\kappa_{max}$ is discussed in the case studies in Section~\ref{sec:Case2}.
%
%
\subsubsection{Entering the HIC region}\label{enterHIC}
As soon as a DW enters the HIC region, two additional processes take place. First, all points surrounding the current operating point are assessed as well, as they may be part of $\Omega$. This is indicated in Fig. \ref{fig_direction} by the yellow circles. Second, we allow the DW to move along only a single dimension (the dimension is still selected based on the steepest descent) and with the minimum step size. This ensures that we collect as many points within the HIC region as possible.
\subsubsection{Termination of the Directed Walks}
Each DW terminates if the next step arrives at an operating point already existing in the database. The termination criterion excludes operating points that were collected as ``surrounding points'' of a current step (see Section~\ref{enterHIC}).
\subsubsection{Full N-1 contingency check}\label{seq:FinalEval}
After all DWs have been performed for every initialization point in parallel, we evaluate all safe (and almost safe) operating points in the database against all possible contingencies to ensure that no violations occur. More formally, we assess all operating points in the final database with $\gamma_k \geq \gamma-\mu$ for all remaining $\left|\mathcal{C} \right|-1$ systems to ensure that a possible false identification of the most critical contingency does not affect the stability boundary detection. This allows us to guarantee a high level of accuracy in determining the security boundary. Despite this being the most computationally expensive step of our method, accounting for over 50\% of the required time, in absence of expert knowledge this procedure is required for any method reported in the literature \cite{Liu2013b,Liu2014}. The difference is, however, that our approach manages to discard a large volume of non-relevant data before this step, and, as a result, outperforms existing methods by being at least 10 to 20 times faster.
%
\subsubsection{Final Set of Directed Walks}\label{seq:Round2}
The maximum number of steps $\kappa_{max}$, although helpful for the efficient parallelization of the DWs, may result in DWs that have not sufficiently explored the search space. In this final step, for any DWs that have reached $\kappa_{max}$ while inside the HIC region, we perform an additional round of DW to explore as thoroughly as possible the HIC region. The final points from the previous round serve as initialization points.

\section{Extension to a N-k Analysis}\label{sec:N-k}
As the authors in \cite{Chen2005} highlight, it is computationally impractical to analyze all N-k contingency sequences, due to the large number of possible contingencies and their combinations. In order to minimize the number of required analyses, different approaches exist in literature to find a subset of plausible harmful N-k contingencies using e.g. time domain simulations \cite{Weckesser2018}, event trees and functional groups \cite{Chen2005} or fault chain theory \cite{Wang2011b}. Each of these methods can be combined with our proposed method to determine in advance the list of plausible N-k contingencies, which can then be used as the set of considered critical contingencies during the Directed Walks.

In this section, however, as we wish to continue with our approach of not requiring any kind of expert knowledge, we extend the security analysis to a N-k scenario. Up to this point, we have identified the HIC region and the security boundary considering N-1 security and small-signal stability (``N-1 and SSS''). Our ultimate goal in this section is to determine how the HIC region and the security boundary should be adjusted if we consider N-k security and small-signal stability. To do that, we start with all \emph{stable} points in the ``N-1 and SSS'' HIC region, as they were identified by the Directed Walks in Section~\ref{sec:directedwalks}. As described in Section~\ref{seq:FinalEval}, a full N-1 contingency assessment is carried out for every final point of the DWs. As a result, we have exact knowledge of the impact of all contingencies, and can rank them from the most critical to the least critical. To extend our analysis to the N-2 case, we apply the two most critical contingencies at the same time to our system. Our goal is to perform Directed Walks from the ``N-1 and SSS'' HIC region to the ``N-2 and SSS'' HIC region, and determine the new security boundary. Admittedly, the combination of the two most critical N-1 contingencies is often but not always the most critical N-2 contingency. The goal of the Directed Walks, however, is to determine a path that will lead towards the new HIC region -- and several combinations of most critical contingencies can lead to that. To ensure that no violations occur, similar to the N-1 case, we perform a N-2 contingency check (along with small-signal stability) at the end of the new Directed Walks. This ensures that all operating points which will land in the database will have been checked if they are N-2 secure for a wide range of contingencies.

Similar is the procedure that can be followed for enforcing N-k security, with k>2. Given the combinatorial nature of the N-k security assessment though, beyond a certain point expert knowledge or advanced methods must be used in order to consider only a limited set of critical N-k contingencies.


The different steps are summarized below taking the \mbox{N-2} security database generation as an example but can be generalized to a N-k case.
\subsection{Initialization Points}
As already mentioned, the initialization points for the ``N-2 and SSS'' security assessment are the final points of the Directed Walks during the ``N-1 and SSS'' procedure described in Section~\ref{sec:directedwalks}. More specifically, it is all \emph{stable} points belonging to the ``N-1 and SSS'' HIC region. Directed Walks often result to OPs close to each other. To cover an as large space as possible keeping the computation time low, we want to pick initialization points that have at least a certain distance between each other. As a result, after picking each initialization point, we assume a radius $R_{N-2}$ around that point, and discard any potential initialization point within this radius. This allows us to have a reduced and more uniformly distributed set of initialization points to start our assessment. The choice of $R_{N-2}$ depends on the maximum number of steps $\kappa_{max,N-2}$ of the directed walks during the N-2 security database generation. As we aim for avoiding duplicates but also for maximizing the number of unique OPs within $\Omega$, a choice of $R_{N-2} \leq \kappa_{max,N-2} \cdot \min\{\alpha_k\}$ is recommended, where $\min\{\alpha_k\}$ is the minimum step size as defined in \eqref{eq:stepsize}.



\subsection{Most Critical N-2 Contingency}
By taking advantage of the full N-1 contingency check, described in section \ref{seq:FinalEval}, we already know the set with the most critical contingencies for every initialization point. The two most critical contingencies for the N-1 case are used as the most critical N-2 contingency for the directed walks during the N-2 security analysis. Please note that the role of the critical contingencies is to determine an appropriate direction of the directed walks towards the new HIC region. Similar to Section~\ref{seq:FinalEval}, a N-2 contingency check (for the chosen N-1 fault) will follow in the end again. This ensures that even if the choice of the most critical N-2 contingency for the directed walks is inaccurate, it will not necessarily result in a falsely classified operating point in the database.

\subsection{Directed Walks}
The directed walks work exactly in the same way as introduced in Section \ref{sec:directedwalks} including a full N-2 contingency check for the given most critical outage (N-1 contingency) in the end. Ideally, a full N-2 contingency check must be carried out for all possible combinations. Given the exponential increase in the number of N-2 contingencies as the system grows larger, expert knowledge or advanced methods become necessary. In our case studies, we did not observe significant changes of the HIC set, while checking for several different pairs of \mbox{N-2} contingencies. However, in other systems, use of advanced methods will be probably necessary to select the set of most critical contingencies to be used for this final check.

\section{Case Studies} \label{sec:Case2}
In the first case study, the efficient database generation method is applied on the IEEE 14 bus system. We measure the efficiency improvement compared with the brute force approach (BF), and we demonstrate how our method outperforms importance sampling techniques. It is impossible to carry out the comparison with the BF approach in larger systems, as BF becomes intractable. In the second case study, we demonstrate the scalability of our method to larger systems, such as the NESTA 162 bus system \cite{Coffrin2014}. In the same case study, we also highlight how the proposed method allows to extend a N-1 security to a N-2 security assessment and emphasize how the high quality of the database generated with our method allow machine learning algorithms to achieve a higher accuracy in the data-driven security assessment. The case studies in this paper use the combination of N-1 (or N-2) security and small-signal stability for the definition of the security boundary. It should be stressed though that the proposed methodology proposes a general framework and is applicable to a number of other stability metrics or power system models.
\subsection{Small-Signal Model}
A sixth order synchronous machine model \cite{Sauer1998} with an Automatic Voltage Regulator (AVR) Type I (3 states) is used in this study. With an additional state for the bus voltage measurement delay this leads to a state-space model of $10\cdot N_G$ states, with $N_G$ representing the number of generators in the grid. In case of the NESTA 162 bus system, all generators are addtionally equipped with Power System Stabilizers (PSS) type 1 adding an additional state per generator. The small signal models were derived using Mathematica, the initialization and small signal analysis were carried out using Matpower 6.0 \cite{Zimmerman2011} and Matlab. Reactive power limits of the generators are enforced. For a detailed description of the derivation of a multi-machine model, the interested reader is referred to \cite{Milano2010}. Machine parameters are taken from \cite{Anderson1977}.
%
\subsection{IEEE 14 bus system}\label{sec:case_study1}
Carrying out the first case study on a small system, where the BF approach is still tractable, allows us to verify that our method is capable of finding $\unit[100]{\%}$ of the points belonging to the HIC region. To ensure comparability, all simulations used 20 cores of the DTU HPC cluster.

Network data is given in \cite{Zimmerman2011}, machine parameters are given in \cite{IREP2017}. The considered contingencies include all line faults (except lines 7-8 and 6-13\footnote{\label{note1}The IEEE 14-bus and the NESTA 162-bus systems, based on the available data, are not N-1 secure for all possible contingencies. The outage of those specific lines lead to violations (e.g. voltage limits, component overloadings, or small-signal instability) that no redispatching measure can mitigate. This would not have happened in a real system. In order not to end up with an empty set of operating points, and still use data publicly available, we choose to neglect these outages.}). 
Due to the BF approach, we know that 1457 operating points belong to the HIC set, i.e. with $\unit[2.75]{\%} < \zeta_{min} < \unit[3.25]{\%}$.
 \begin{table}[!t]
 \vspace{-3ex}
\caption{RESULTS: IEEE 14 BUS SYSTEM
  }
\label{tab:resultsIEEE14}
\centering
\begin{tabular}{rrrcl}
\hline
\centering Required  & Time in \%&\centering OPs in $\Omega$ &Method&\\
\centering time&  w.r.t. BF  & \centering found & & $\eta_1 \; /\; \eta_2 \; /\; \eta_3 \; /\; \kappa_{max}$\\
\hline
$\unit[2.56]{min}$&$\unit[0.46]{\%}$&$\unit[95.13]{\%}$& DWs &$0\; /\;200\; /\; 2k\; /\; 10$\\
$\unit[2.99]{min}$&$\unit[0.54]{\%}$&$\unit[98.9]{\%}$&DWs &$0\; /\;200\; /\; 2k\; /\; 15$\\
$\unit[2.94]{min}$&$\unit[0.53]{\%}$&$\unit[97.80]{\%}$& DWs &$0\; /\;200\; /\; 2k\; /\; 20$\\
\cellcolor{green!25}$\unit[3.77]{min}$&\cellcolor{green!25}$\unit[0.68]{\%}$&\cellcolor{green!25}$\unit[100]{\%}$& \cellcolor{green!25}DWs &\cellcolor{green!25}$ 0\; /\; 200\; /\; 2k\; /\; 25$\\
$\unit[2.94]{min}$&$\unit[0.53]{\%}$&$\unit[97.80]{\%}$& DWs &$0\; /\;200\; /\; 1k\; /\; 20$\\
$\unit[3.48]{min}$&$\unit[0.74]{\%}$&$\unit[99.93]{\%}$& DWs &$ 0\; /\; 200\; /\; 3k\; /\; 20$\\
$\unit[4.80]{min}$&$\unit[0.86]{\%}$&$\unit[100]{\%}$& DWs &$ 0\; /\; 200\; /\; 5k\; /\; 20$\\
\hline
$\unit[37.0]{min}$&$\unit[6.66]{\%}$&$\unit[100]{\%}$& \multicolumn{2}{l}{Importance Sampling (IS)} \\
\hline
$\unit[556]{min}$&$\unit[100]{\%}$&$\unit[100]{\%}$& \multicolumn{2}{l}{Brute Force (BF)}
\end{tabular}
\vspace{-3ex}
\end{table}
The grid pruning without considering any contingency does not reduce the search space in this case study; this is because all possible combinations of generation setpoints do not violate any limits for the given load profile. 
Thus, $\eta_1$ is chosen as 0 and we directly start with the contingency-constrained grid pruning considering the five most critical contingencies. 
Table \ref{tab:resultsIEEE14} compares the performance of our method with the BF approach and an Importance Sampling (IS) approach \cite{Liu2013b}. Our method is capable of creating a database including \emph{all} points of interest in \unit[3.77]{min}; that is \unit[0.68]{\%} of the time required by the BF approach (\unit[9.26]{hours}; 147 times faster). 
The proposed method is also significantly faster (approx. 10 times) than an Importance Sampling approach (\unit[37.0]{min}).

One of the major advantages of our method is the drastic search space reduction through the grid pruning and the most critical contingency identification. In this case study, grid pruning eliminated up to \unit[70.13]{\%} of all $\approx \unit[2.5 \cdot 10^6]{}$ potential operating points (the number varies based on the number of initialization points). At the same time, performing every DW step for the single most critical contingency, we reduce the required assessment from $\left|\mathcal{C} \right|$ systems to 1 system. In larger systems the speed benefits will be even more pronounced, e.g. 14-bus: $\left|\mathcal{C} \right| = 19$ contingencies are reduced to 1 (most critical); 162-bus: $\left|\mathcal{C} \right| = 160$ contingencies reduced to 1.
%

Table \ref{tab:resultsIEEE14} also compares the method's performance for different numbers of initialization points $\eta_{1-3}$ and maximum number of DW steps $\kappa_{max}$.
In this system, choosing a higher number of maximum steps instead of a higher number of initialization points leads to time savings. The same holds in larger systems, as shown in Table~\ref{tab:results162Bus}.

In the highlighted case of Table~\ref{tab:resultsIEEE14}, the required computation time for the different parts of our method is split as follows: \unit[26.67]{\%} (\unit[60.31]{s}) for the grid pruning considering the 5 most critical contingencies (200 operating points); \unit[53.1]{\%} (\unit[120.12]{s}) for the Directed Walks; and \unit[20.24]{\%} (\unit[45.78]{s}) for the final N-1 check of all operating points. Grid pruning eliminates 1149 from the $\eta_3 = 2000$ initialization points, resulting in 851 feasible starting points for the DWs. The most critical contingency is detected correctly in \unit[94.55]{\%} of cases.

\subsection{NESTA 162 bus system}\label{sec:case_study2}
In the second part of the case study, we demonstrate and compare the performance of our method with an Importance Sampling (IS) approach for N-1 security assessment of the NESTA 162 Bus system. A BF approach with a $\unit[1]{MW}$ step size for this system requires the assessment of $7.6 \cdot 10^{29}$ operating points for a single load profile. The assessment of all those points becomes computationally intractable. Thus, the absolute number of unique OPs in $\Omega$ is unknown for this system. Therefore, we focus on highlighting that the proposed method finds significantly more unique OPs close to the security boundary, i.e. creates a database of higher quality, in comparable time frames. Then, we demonstrate how the higher quality of the database allows machine learning algorithms to achieve a higher accuracy within a data-driven security assessment. Finally, we demonstrate how the proposed method allows to extend the N-1 security assessment to a N-k security assessment as described in section \ref{sec:N-k}, here focusing on N-2.
\subsubsection{Database Generation for N-1 Security and Small-Signal Stability Assessment}
The set of considered contingencies includes 159 line faults\footnotemark[1]. 
To ensure comparability, all simulations for the 162-bus system have been performed using 80 cores of the DTU HPC cluster.
\begin{table}[!t]
\caption{RESULTS: NESTA 162 BUS SYSTEM}
\label{tab:results162Bus}
\centering
\begin{tabular}{rrrc}
\hline
\centering Req.  & Unique & \centering Method&\\
 \centering time&OPs in $\Omega$ & & $\eta_1 \; /\; \eta_2 \; /\; \eta_3 \; /\; \kappa_{max}$\\
\hline
$\unit[9.35]{h}$&$3118$& Directed Walks &$30k\; /\;120k\; /\; 800k\; /\; 5$\\
$\unit[13.17]{h}$&$4166$&Directed Walks &$30k\; /\;120k\; /\; 800k\; /\; 10$\\
$\unit[14.57]{h}$&$25046$&Directed Walks &$30k\; /\;120k\; /\; 800k\; /\; 20$\\
$\unit[29.78]{h}$&$150790$& Directed Walks &$ 30k\; /\; 120k\; /\; 800k\; /\; 30$\\
\cellcolor{green!25}$\unit[37.07]{h}$&\cellcolor{green!25}$183295$& \cellcolor{green!25}Directed Walks &\cellcolor{green!25}$ 30k\; /\; 120k\; /\; 800k\; /\; 40$\\
$\unit[13.36]{h}$&$16587$& Directed Walks &$ 100k\; /\;200k\; /\; 800k\; /\; 5$\\
$\unit[18.20]{h}$&$45040$& Directed Walks &$ 100k\; /\;200k\; /\; 800k\; /\; 10$\\
\hline
$\unit[35.70]{h}$& 901& \multicolumn{2}{l}{Importance Sampling (IS)}
\end{tabular}
\vspace{-3ex}
\end{table}
Compared to the IEEE 14 bus system, the problem size (potential \# of OPs) is 23 orders of magnitude larger while the problem complexity (\# of faults) increased 6.2 times. Table~\ref{tab:results162Bus} presents the results of our method compared with an Importance Sampling approach \cite{Liu2013b}. As the BF approach for this system is intractable, the exact number of points within the HIC region (set $\Omega$) is unknown. Therefore, the focus here is on demonstrating that within similar time frames, our proposed method is capable of finding substantially more unique operating points inside $\Omega$. Indeed, our approach identifies approx. three orders of magnitude more HIC points than an Importance Sampling approach (183'295 vs 901 points).

In the highlighted case of Table~\ref{tab:results162Bus}, the computation time is split as follows: $\unit[3.44]{h}$ ($\unit[9.28]{\%}$) for LHS (3 stages), $\unit[1.85]{h}$ ($\unit[4.98]{\%}$) for both stages of grid pruning, $\unit[7.04]{h}$ ($\unit[18.98]{\%}$) for the DWs, and $\unit[24.75]{h}$ ($\unit[66.76]{\%}$)
for the final N-1 check of all operating points of interest. This highlights that the most computationally expensive part is the complete N-1 analysis and shows why our proposed method is significantly faster than others: (i) we reduce the search space by eliminating infeasible N-1 points through the grid pruning algorithm, (ii) we evaluate most points only for one contingency and discard all with $\zeta < \unit[2.75]{\%}$, and (iii) the method can largely be scheduled in parallel.
\subsubsection{N-1 Security and Small Signal Stability Assessment}
As the topic of this paper is the efficient database generation for a data-driven security assessment, we briefly want to highlight how important the higher quality of the databases created by the proposed method is for such a data-driven security assessment. There are three important factors to be considered here: (i) The more time is needed to create a database of high quality, the longer is the wait before the machine learning algorithm can start its training phase. (ii) The more data is needed to sufficiently describe a system, the longer the algorithm needs to be trained. (iii) The further away points are from the security boundary, the less information they contain about the security boundary. Thus, it is essential that the database contains many unique points close to the security boundary enabling the algorithm to determine the security boundary as accurately as possible \cite{Krishnan2011}.

In order to demonstrate the impact of the higher quality of the database on the achievable accuracy of machine learning algorithms, we implement a data-driven N-1 security and small-signal stability assessment using a decision tree. Based on lessons learned from previous works \cite{IREP2017}, we use the active and reactive power flows on the lines as predictors and let the decision tree classify between `fulfilling the requirements' and `not fulfilling the requirements' i.e. N-1 secure \emph{and} a minimum damping ratio $\zeta_{min} \geq \unit[3]{\%}$, or not. This decision tree then could be included in an optimal power flow, or in general an optimization framework, as shown in  \cite{IREP2017,Lejla_PSCC}. 

We compare the highlighted database in Table~\ref{tab:results162Bus} with the one created with Importance Sampling and also presented in analyzed in Table \ref{tab:results162Bus}. Both required a comparable computation time to be generated. For a fair comparison, we use as training set all assessed operating points in each case; i.e. not only the unique OPs in the HIC region listed in Table \ref{tab:results162Bus}, but rather all OPs that were found in the safe, unsafe, and HIC regions during these 35-37 hours. To simplify the comparison, we used Matlab to train a simple decision tree using the standard Classification and Regression Tree (CART) algorithm with Gini's diversity index as splitting criterion and without limiting the tree depth. In order to avoid over-fitting, we used Matlab to apply cost-complexity pruning, minimizing the cross-validated classification error of the trees.

To test the accuracy of the two decision trees, one of which was trained on the database created with our approach and the other a database created with the Importance Sampling method, we use a common test set of 90'000 operating points. To avoid favoring one of the two database generation methods, we created the test set by merging two datasets. Thus, $50\%$, i.e. 45'000 data points, are generated through an importance sampling approach using initialization values different from the ones used for the training set. The other 45'000 data points of the test set are taken from our last database generation attempt with our method shown in Table~\ref{tab:results162Bus}, where we generated 45'040 unique points in $\Omega$ in 18.20 h (please note though that the 45'000 points of the test set were picked from a wide range of points generated during that process, which are located in the safe, unsafe, and HIC region). For that attempt we used $\eta_1 = 100k$, $\eta_2 = 200k$ and $\eta_3 = 800k$, which means that this part of the test set also had different initialization points from the training set. In both cases we ensured that none of the data-points in the test set were part of any training set. Thus, this is completely unseen data and allows to evaluate the generalization capability of the trained classifiers.

The decision tree trained on the highlighted database in Table \ref{tab:results162Bus}, which was generated with our method, achieves an accuracy of $\unit[85.91]{\%}$ while the tree trained on the database created with IS achieves an accuracy of $\unit[73.00]{\%}$.
As properties of the test set might have an impact on the accuracy score, additional measures are usually examined for the performance of machine learning algorithms. Besides accuracy, an important measure of the quality of the classification is the number of true and false positives and negatives. The Matthews correlation coefficient (MCC) is generally regarded as a balanced measure for the quality of the binary classification. The MCC\footnote{\label{note2}$MCC=\frac{TP\cdot TN - FP \cdot FN}{\sqrt{(TP+FP)(TP+FN)(TN+FP)(TN+FN)}}, \quad -1\leq MCC \leq 1$ with $TP$ and $TN$ representing the correctly identified positive and negative samples and $FP$ and $FN$ representing the falsely classified negative and positive samples \cite{Chicco2017}.} is in essence a correlation coefficient between the observed and predicted binary classifications, returning a value between -1 and +1. A value of +1 means perfect prediction, 0 means no better than random prediction, and -1 indicates total disagreement. In our case, the decision tree trained with our method has $MCC_{DW}=0.6247$, while the tree trained with importance sampling has $MCC_{IS}=-0.0943$. This highlights the over-optimistic results of the accuracy measure and the significant better performance of the tree trained on the database created with our proposed method. 

As the only difference between the configuration of the two decision trees is the database each tree was trained on, this emphasizes the importance of a high quality database to achieve the best possible results in a data-driven security assessment. More advanced machine learning algorithms may achieve even better results; this is, however, out of the scope of this paper. All created operating points are collected and published on GitHub \cite{GitHubDatabase}.


\subsubsection{Database Generation for N-2 Security Assessment}
To demonstrate how the proposed method is capable of extending a N-1 security assessment to an N-k security assessment, we used the highlighted N-1 database in Table \ref{tab:results162Bus} as a starting point. Similar to the N-1 case, for specific contingencies we were unable to obtain a N-2 secure system\footnotemark[1]. As a result, we had to relax the voltage limits to $\unit[0.9]{p.u.} \leq V_i \leq \unit[1.1]{p.u.}$ for all contingencies, and remove the line fault on the line between bus 125 and bus 126 from the contingency list. Thus, in the \mbox{N-2} security analysis the set of considered contingencies includes 158 line faults.

In order to avoid the creation of unnecessary duplicates and minimize computation, all data-points located in the vicinity of other data-points from the HIC region of the N-1 security analysis, i.e. all OPs located within a radius of $\unit[5]{MW}$ surrounding another OP, are discarded. The remaining OPs serve as initialization points for a new round of directed walks as described in section \ref{sec:N-k}. Within $\unit[8.5]{h}$ we obtain a database with 52'107 unique OPs that belong to $\Omega_{N-2}$. $\unit[33.2]{\%}$ of the time, i.e. $\unit[2.8]{h}$, is used for the directed walks while $\unit[66.8]{\%}$ of the time, i.e. $\unit[5.7]{h}$, is required by the final N-2 contingency check. Hence, we obtain a $\unit[24]{\%}$ speed-up compared to the creation of the N-1 security database highlighted in Tab. \ref{tab:results162Bus}. This speed-up is achieved because the first half of the method, i.e. the creation of the initialization points and the grid pruning, is not required when starting from the N-1 case.

\section{Conclusions} \label{sec:conclusion}
This work proposes an efficient database generation method that can accurately determine power system security boundaries, while drastically reducing computation time. Such databases are fundamental to any Dynamic Security Assessment (DSA) method, as the information in historical data is not sufficient, containing very few abnormal situations. This topic has not received the appropriate attention in the literature, with the few existing approaches proposing methods based on importance sampling. 

Our approach is highly scalable, modular, and achieves drastic speed-ups compared with existing methods. It is composed of two parts. First, the search space reduction, which quickly discards large infeasible regions leveraging advancements in convex relaxation techniques and complex network theory approaches. Second, the ``Directed Walks'', a highly parallelizable algorithm, which efficiently explores the search space and can determine the security boundary with extremely high accuracy. Using a number of initialization points, a variable step size, and based on a steepest descent method, the ``Directed Walk'' algorithm traverses fast through large parts of feasible (or infeasible) regions, while it focuses on the high information content area in the vicinity of the security boundary. Our case studies on the IEEE 14-bus and the NESTA 162-bus system demonstrate the high quality, high scalability and excellent performance of our algorithm. They are able to identify up to 100\% of the operating points around the security boundary, while achieving computational speed-ups of over 10 to 20 times compared with an importance sampling approach. We also demonstrated the importance of a high quality database to achieve the best possible results in a data-driven security assessment. Given equal computation time, training machine learning algorithms with the database generated by our method clearly outperforms other approaches.

Our approach is modular, not dependent on the initial sampling set (as importance sampling is), and agnostic to the security criteria used to define the security boundary. Criteria to be used include N-1 or N-k security, small-signal stability, voltage stability, or a combination of several of them. The method can find application in off-line security assessment, in real-time operation, and in machine learning and other data-driven applications, providing a computationally efficient way to generate the required data for training and testing of new methods.
\vspace{-0.2cm}
\section{Acknowledgements}
\footnotesize This work has been supported by  the  EU-FP7 Project ``Best Paths'', grant agreement no.~612748, and by the Danish ForskEL project ``Best Paths for DK'', grant agreement 12264.
\vspace{-0.2cm}
\bibliographystyle{IEEEtran}
\bibliography{library}
\vspace{-3ex}
\begin{IEEEbiography}[{\includegraphics[width=1in,height=1.25in,clip,keepaspectratio]{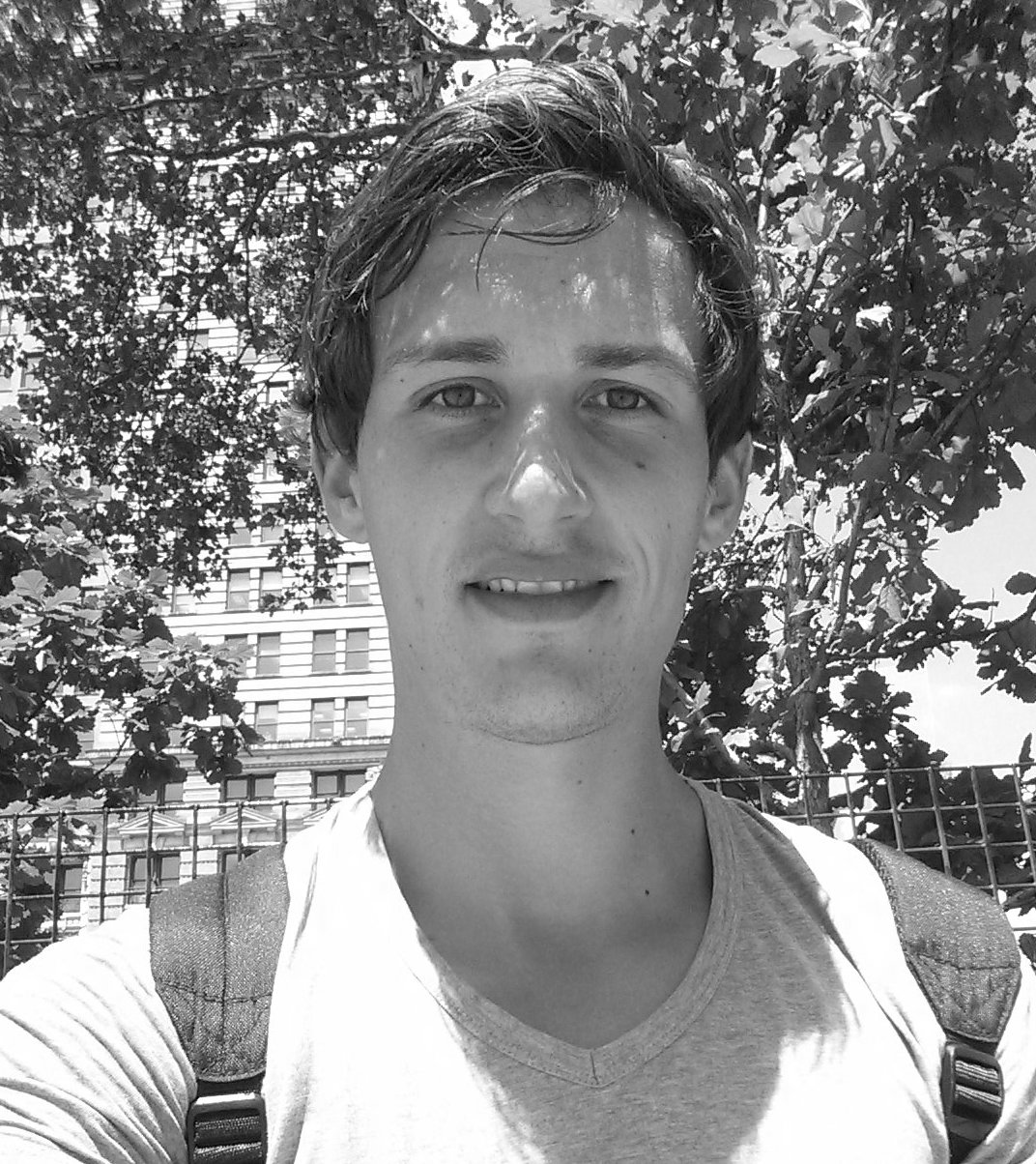}}]{Florian Thams} (S' 15)
received the B.Sc. and M.Sc. in Business Administration and Engineering from the Christan Albrechts University of Kiel, Germany, in 2012 and 2015, respectively. From February '15 to June '15 he joined the Center of Electric Power and Energy at the Technical University of Denmark (DTU) as Research Assistant. Currently he is pursuing his Ph.D. at the same institute. His research interests include power system dynamics and stability, HVDC systems, and machine learning applications.
\end{IEEEbiography}
\begin{IEEEbiography}[{\includegraphics[width=1in,height=1.25in,clip,keepaspectratio]{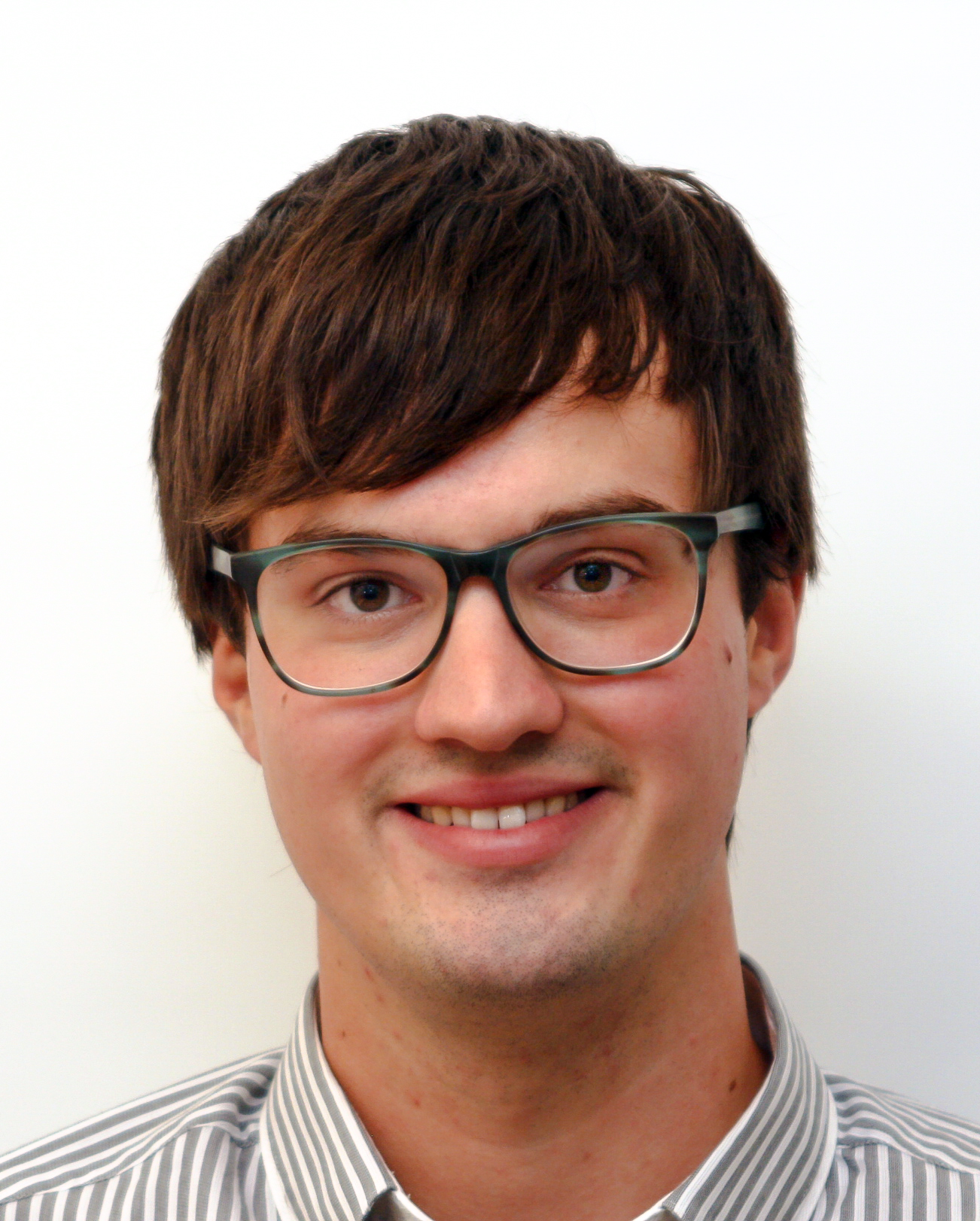}}]{Andreas Venzke} (S' 16) received the M.Sc. degree in Energy Science and Technology from ETH Zurich, Zurich, Switzerland in 2017. He is currently working towards the Ph.D. degree at the Department of Electrical Engineering, Technical University of Denmark (DTU), Kongens Lyngby, Denmark. His research interests include power system operation under uncertainty and convex relaxations of optimal power flow.
\end{IEEEbiography}
\begin{IEEEbiography}[{\includegraphics[width=1in,height=1.25in,clip,keepaspectratio]{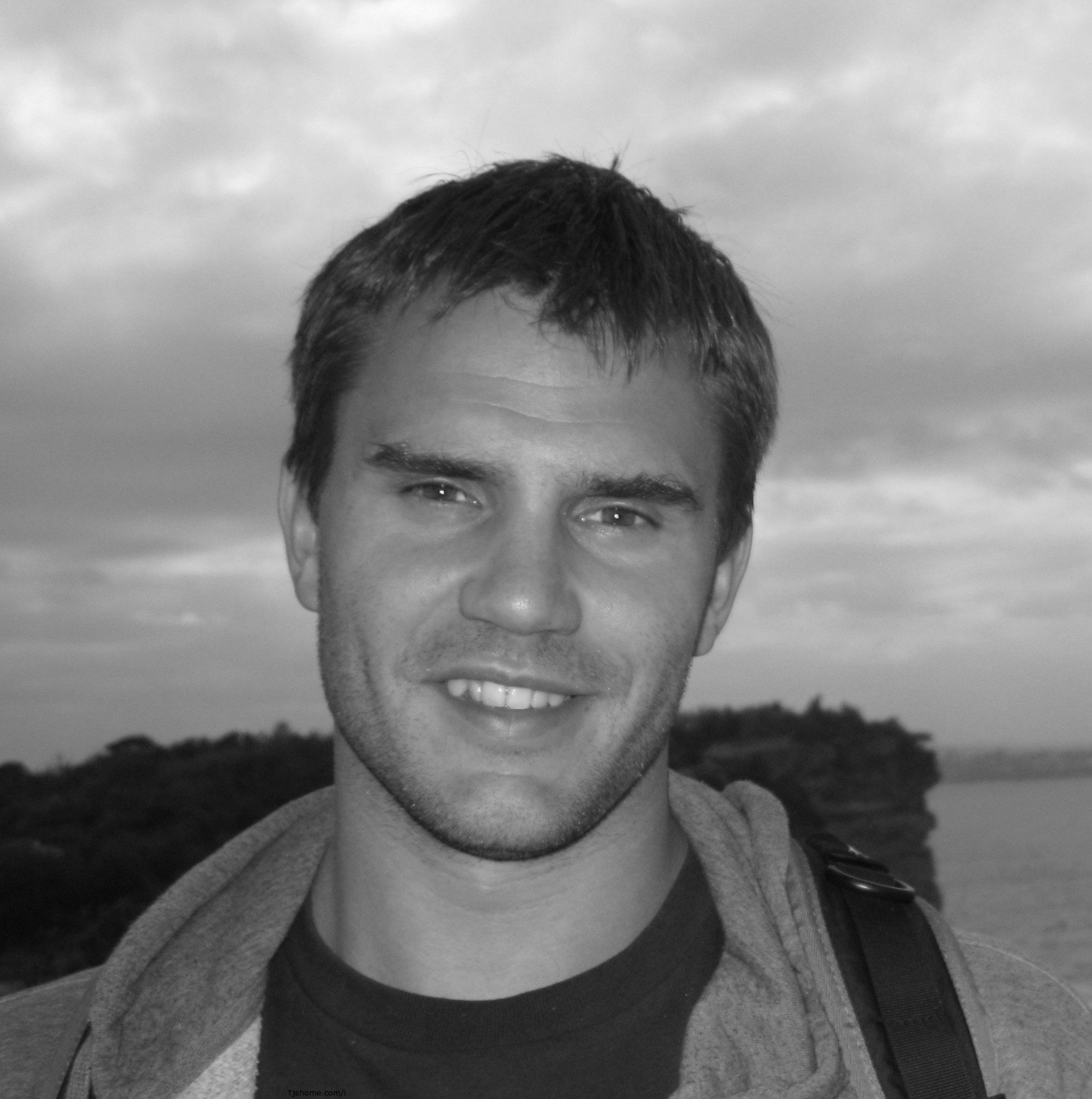}}]{Robert Eriksson} (S'06, M'11, SM'16) received his M.Sc. and Ph.D. degrees in electrical engineering from the KTH Royal Institute of Technology, Stockholm, Sweden, in 2005 and 2011, respectively. Between 2013 and 2015, he was an Associate Professor with the Center for Electric Power and Energy (CEE), DTU Technical University of Denmark. He is currently with the Swedish National Grid, Sundbyberg, Sweden, involved in research and development at the Department of Market and System Development. He is also an affiliated faculty with the KTH Royal Institute of Technology, Stockholm, Sweden. His research interests include power system dynamics and stability, HVDC systems, dc grids, and automatic control. In 2018, he received the title of Docent at the KTH Royal Institute of Technology.
\end{IEEEbiography}
\begin{IEEEbiography}[{\includegraphics[width=1in,height=1.25in,clip,keepaspectratio]{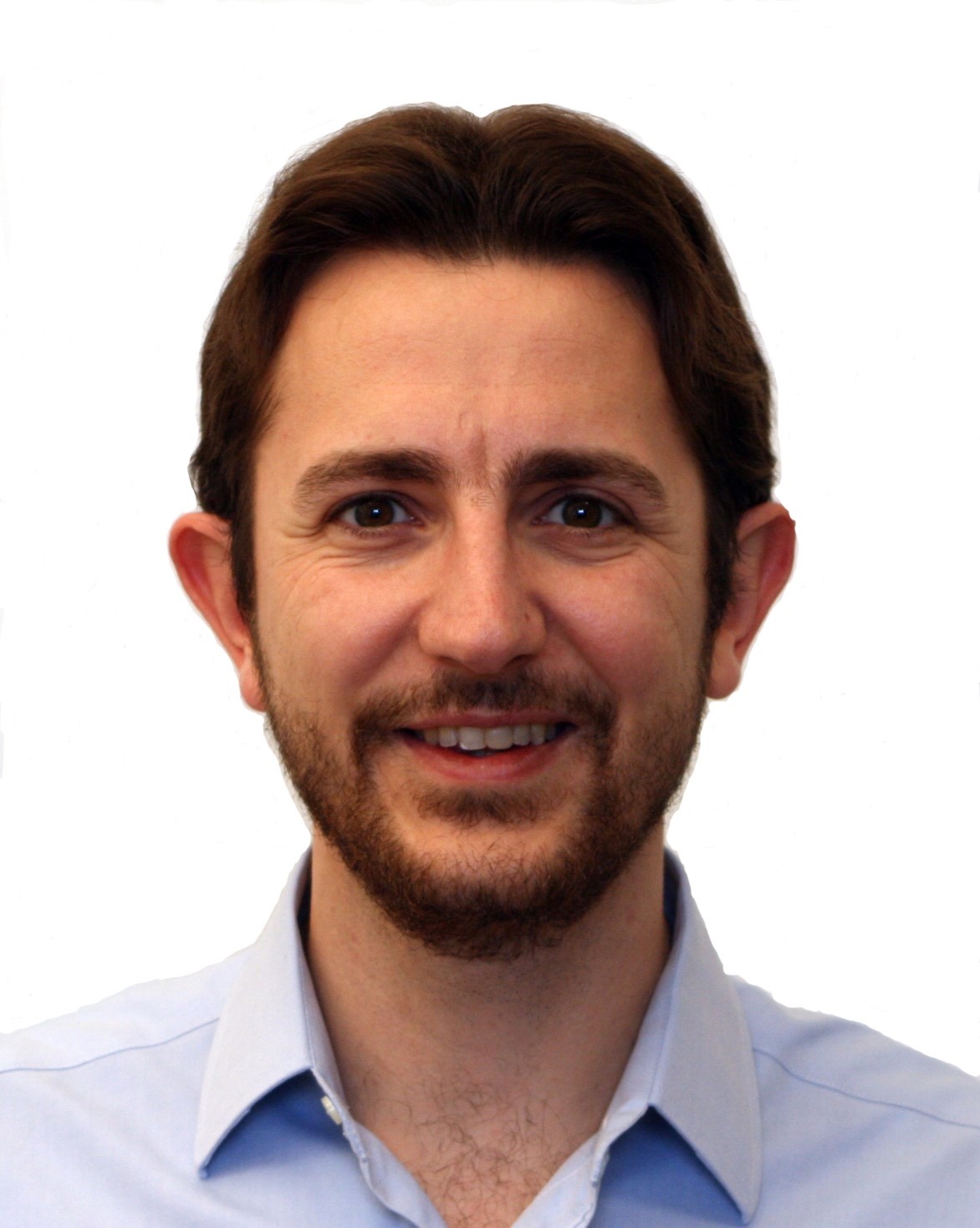}}]{Spyros Chatzivasileiadis} (S'04, M'14, SM'18) is an Associate Professor at the Technical University of Denmark (DTU). Before that he was a post- doctoral researcher at the Massachusetts Institute of Technology (MIT), USA and at Lawrence Berkeley National Laboratory, USA. Spyros holds a PhD from ETH Zurich, Switzerland (2013) and a Diploma in Electrical and Computer Engineering from the National Technical University of Athens (NTUA), Greece (2007). In March 2016 he joined the Center of Electric Power and Energy at DTU. He is currently working on power system optimization and control of AC and HVDC grids, including semidefinite relaxations, distributed optimization, and data-driven stability assessment.
\end{IEEEbiography}
%
 \end{document}